\documentclass[twocolumn,twoside,nofootinbib]{revtex4}
\usepackage{graphicx}
\usepackage{fancyhdr}
\preprint{TTP--19--015}

\newcommand{\rd}{\ensuremath{{\cal R}(D)}\;}
\newcommand{\rds}{\ensuremath{{\cal R}(D^*)}\;}
\newcommand{\rdrds}{{\cal R}(D^{(*)})}

\newcommand{\bbc}{\text{BR}(B_c\to \tau\nu_\tau)}

\newcommand{\rlam}{{\cal R}(\Lambda_c)}

\newcommand{\bdstau}{\ensuremath{B\to D^*\tau\nu\;}}

\newcommand{\ptauds}{\ensuremath{P_\tau(D^\ast)}}
\newcommand{\fld}{\ensuremath{F_{L}(D^{\ast})}}

\newcommand{\dline}[1]{{\parbox{3.5em}{\rule{0cm}{2.3ex}{#1}\strut}}}
\newcommand{\dlinem}[1]{{\parbox{2.5em}{\rule{0cm}{2.3ex}{#1}\strut}}}

\usepackage{comment}
\usepackage{hyperref}

\begin{document}

\title{New Physics in $\boldmath{b\to c \tau \nu}$: Impact of Polarisation Observables and $\boldmath{B_c\to\tau\nu}$}

%
\author{Marta Moscati}
\email{marta.moscati@kit.edu}
\affiliation{Institut f\"ur Theoretische Teilchenphysik (TTP), Karlsruher Institut f\"ur Technologie (KIT), 76131 Karlsruhe, Germany}
\begin{abstract}
The experimental values of the lepton-flavour-universality tests $\rd$ and $\rds$ show a tension of about $3.1\sigma$ with their Standard Model prediction. Motivated by this tension, we perform a fit of the $b\to c\tau\nu$ data. We consider one-particle scenarios imposing consecutive limits on $\bbc$, and analyse how these limits affect the fits. We include the polarisation observables available to date and predict those that are still to be measured, and conclude that they have a high model-resolving power. For each scenario we also predict $\rlam$, observing that an enhancement of $\rdrds$ implies an enhancement of $\rlam$ in any scenario. We trace back this enhancement to a sum-rule valid irrespective of the scenario used to fit $\rdrds$.

\end{abstract}

\maketitle

\thispagestyle{fancy}


\section{Introduction}
\begin{figure}[h]
	\centering
	\includegraphics[width=0.48\textwidth,bb= 0 0 511 331]{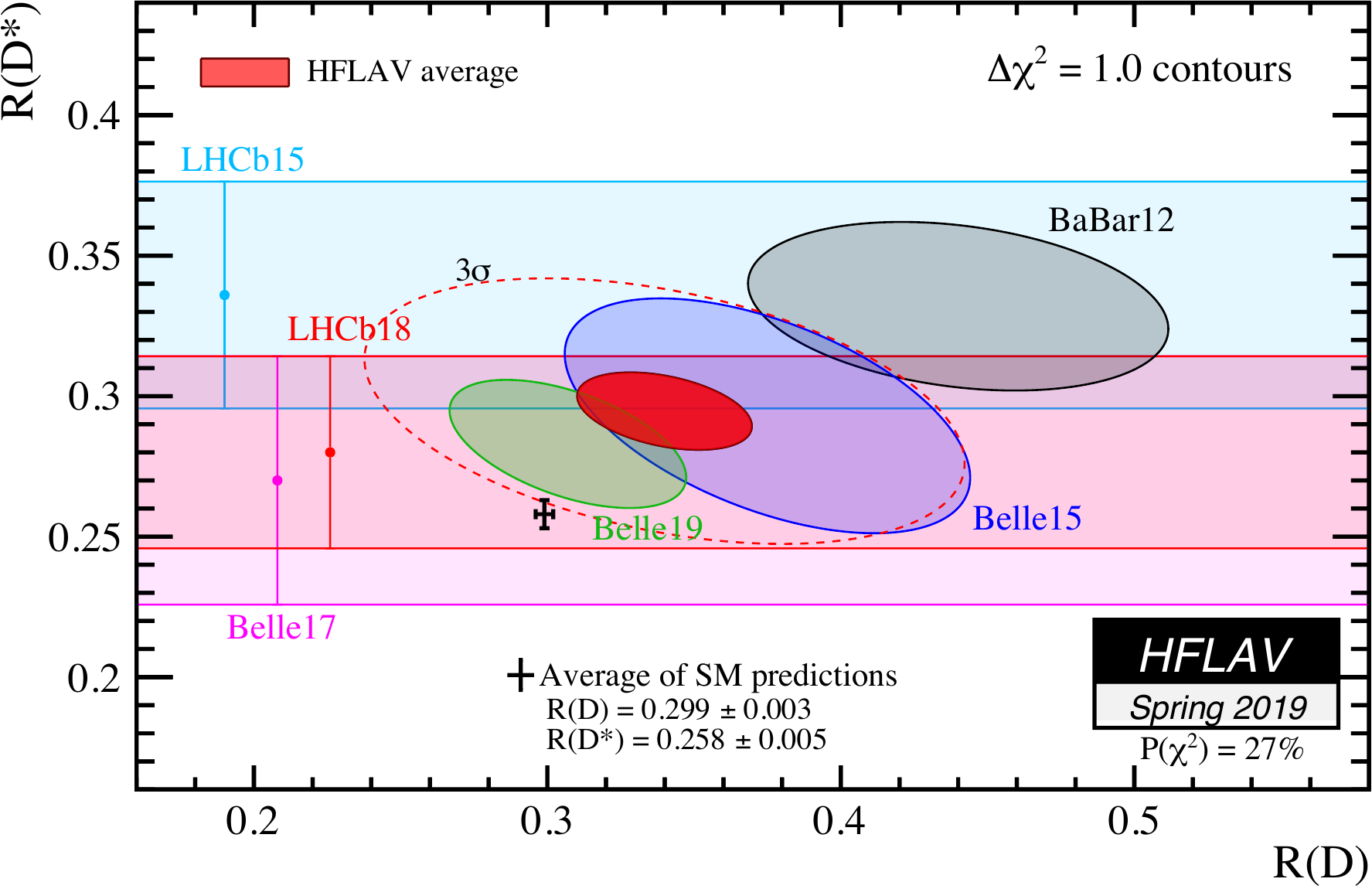}
\caption{Summary plot of the measurements of $\rdrds$, taken from~\cite{Amhis:2016xyh}.}
\label{Fig:ellipse}
\end{figure}

The lepton-flavour-universality tests $\rdrds \equiv{\rm BR}(B\to D^{(*)} \tau\nu)/{\rm BR}(B\to D^{(*)} \ell \nu)$, measured by the BaBar, Belle and LHCb collaborations~\cite{Lees:2012xj,Lees:2013uzd, Huschle:2015rga,Sato:2016svk,Hirose:2016wfn,Hirose:2017dxl,Aaij:2015yra,Aaij:2017uff,Aaij:2017deq,abd:2019dgh}, are in tension with the Standard Model (SM) prediction with a combined difference of about $3.1\sigma$. The average of the measurements can be found in~\cite{Amhis:2016xyh}, and Figure~\ref{Fig:ellipse} displays a summary plot.
Data on the angular distribution of the final state particles in $\bdstau$ are also available from the Belle collaboration~\cite{Hirose:2016wfn,Hirose:2017dxl,Adamczyk}

\begin{equation}
\begin{array}{rcl}
	F_{L}(D^{\ast})&=&\frac{\Gamma(B \to D^{*}_L \tau \nu)}{\Gamma(B \to
  D^{*} \tau \nu)}\nonumber\\
  	&=& 0.60\pm 0.08 \pm 0.035\nonumber\\
  	P_\tau(D^\ast) &=&\frac{\Gamma(B \to D^{*} \tau^{\lambda=+1/2} \nu)-\Gamma(B \to D^{*} \tau^{\lambda=-1/2} \nu)}{\Gamma(B \to
  D^{*} \tau \nu)}\nonumber
  	\\&=&-0.38\pm 0.51^{+0.21}_{-0.16}\nonumber
\label{eq:pol}
\end{array}
\end{equation}
In our analysis~\cite{Blanke:2018yud,Blanke:2019qrx} we fitted these data to scenarios of new physics (NP) in which a single heavy mediator contributes to the transition $b\to c \tau\nu$, without contributing to the channels with a light lepton\footnote{For an analysis of NP effects in $b\to u\ell\nu$ see~\cite{Colangelo:2019axi}.}.

\section{New physics scenarios}
The contributions  of a NP mediator with mass above the $B$ meson mass to $b\to c\tau\nu$ transitions, excluding the presence of light right-handed neutrinos, can be parametrised in terms of an effective field theory (EFT) as
\begin{equation}
\renewcommand{\arraystretch}{1.8}
\begin{array}{r}
 {\cal H}_{\rm eff}=  2\sqrt{2} G_{F} V^{}_{cb} \big[(1+C_{V}^{L}) O_{V}^L +   C_{S}^{R} O_{S}^{R} 
 \\   +C_{S}^{L} O_{S}^L+   C_{T} O_{T}\big] \,,\quad\quad
\end{array}
\label{Heff}
\end{equation}
with
\begin{equation}
\begin{array}{rcl}
   O_{V}^L  &=& \left(\bar c\gamma ^{\mu } P_L b\right)  \left(\bar\tau \gamma_{\mu } P_L \nu_{\tau}\right), \\ 
   O_{S}^R  &=& \left( \bar c P_R b \right) \left( \bar\tau P_L \nu_{\tau}\right), \\
   O_{S}^L  &=& \left( \bar c P_L b \right) \left( \bar\tau P_L \nu_{\tau}\right),   \\
   O_{T}  &=& \left( \bar c \sigma^{\mu\nu}P_L  b \right) \left( \bar\tau \sigma_{\mu\nu} P_L \nu_{\tau}\right).   \\
\end{array}
\label{Oeff}
\end{equation}
The $1$ in the vectorial coupling represents the SM contribution, while all the remaining Wilson coefficients (WCs) encode only NP contributions.

The addition of a single NP particle to the SM can only give rise to a restricted subset of combinations of WCs. With the further assumption of real couplings, the parameters to fit are at most two. We can hence have the one-dimensional scenarios\footnote{For a discussion of the effects of a tensor coupling see~\cite{Biancofiore:2013ki,Colangelo:2016ymy,Colangelo:2018cnj}.}: 
\begin{itemize}
\item {\boldmath $C_V^L$}: arising from the SU(2$)_L$-singlet vector leptoquark (LQ) $U_1$~\cite{Alonso:2015sja,Calibbi:2015kma,Fajfer:2015ycq,Barbieri:2015yvd,Barbieri:2016las,Hiller:2016kry,Bhattacharya:2016mcc,Buttazzo:2017ixm,Kumar:2018kmr,Assad:2017iib,DiLuzio:2017vat,Calibbi:2017qbu,Bordone:2017bld,Barbieri:2017tuq,Blanke:2018sro,Greljo:2018tuh,Bordone:2018nbg,Matsuzaki:2018jui,Crivellin:2018yvo,DiLuzio:2018zxy,Biswas:2018snp},
  the scalar SU(2$)_L$-triplet and/or scalar SU(2$)_L$-singlet LQ~\cite{Deshpande:2012rr,Tanaka:2012nw,Sakaki:2013bfa,Freytsis:2015qca,Bauer:2015knc,Cai:2017wry,Crivellin:2017zlb,Altmannshofer:2017poe,Marzocca:2018wcf} with left-handed couplings only, or in models with left-handed $W^\prime$
  bosons~\cite{He:2012zp,Greljo:2015mma,Boucenna:2016wpr,He:2017bft}.
\item {\boldmath $C_S^R$}: arising from charged scalars or from the SU(2)$_L$-doublet vector LQ $V_2$~\cite{Kosnik:2012dj,Biswas:2018iak}.
\item \begin{boldmath}$C_S^L$\end{boldmath}: arising from charged scalars in the hypothesis of a mechanism making $O_S^L$ the dominant
operator~\cite{Crivellin:2012ye,Crivellin:2013wna,Celis:2012dk,Ko:2012sv,Crivellin:2015hha,Dhargyal:2016eri,Chen:2017eby,Iguro:2017ysu,Martinez:2018ynq,Biswas:2018jun}.
\item \begin{boldmath}$C_S^L=4C_T$\end{boldmath}: arising from the scalar SU(2$)_L$-doublet $S_2$ (also called
  $R_2$) LQ~\cite{Becirevic:2016yqi,Becirevic:2018afm}. Note that the relation holds at the NP scale, and gets modified by QCD and electroweak (EW)
  renormalization-group (RG) effects~\cite{Alonso:2013hga,Gonzalez-Alonso:2017iyc}.
\end{itemize}
or the two-dimensional scenarios
\begin{itemize}
\item {\boldmath $(C_V^L,\,C_S^L=-4C_T)$}: arising from the SU(2$)_L$-singlet scalar LQ ($S_1$). The relation $C_S^L=-4C_T$ holds again at the NP scale and must be evolved to the $m_B$ scale~\cite{Gonzalez-Alonso:2017iyc}.
\item {\boldmath $(C_S^R,\,C_S^L)$}: arising from charged scalars.
\item {\boldmath $(C_V^L,\,C_S^R)$}: arising from vector LQs like the SU(2$)_L$-singlet LQ $U_1$.
\item  {\boldmath $({\rm Re}[C_S^L=4C_T],\,{\rm Im}[C_S^L=4C_T])$}: as pointed out in~\cite{Becirevic:2018afm}, the scenario $C_S^L=4C_T$ is able to reproduce the $\mathcal{R}({D^{(*)}})$ data only under the assumption of complex couplings. For this reason we also include it in the two-dimensional fits, fitting separately the real and the imaginary part.
\end{itemize}

\section{Constraints from $\boldmath{\bbc}$}
The vector ($C_V^L$) and pseudoscalar ($C_P = C_S^R-C_S^L$) couplings also mediate the decay $B_c\to \tau\nu$~\cite{Gonzalez-Alonso:2016etj,Alonso:2016oyd}. Although the branching ratio $\bbc$ has not been measured yet, the comparison between the measured and SM-expected~\cite{Gershtein:1994jw,Bigi:1995fs,Beneke:1996xe,Chang:2000ac,Kiselev:2000pp} $B_c$ lifetime allows to set an upper limit on $\bbc$. This approach was used in~\cite{Alonso:2016oyd} to set an upper limit of $30\%$. This limit can be relaxed if one takes into account the uncertainties in the theoretical calculation of the lifetime, originating from the large dependance on $m_c$ and from the calculation methods applied, namely heavy quark expansion and non-relativistic QCD (NRQCD). 

Furthermore, the authors of~\cite{Akeroyd:2017mhr} set the upper limit $\bbc<10\%$ using LEP data from an admixture of $B_c\to\tau\nu$ and $B^-\to\tau\nu$ and using the fragmentation functions ratio $f_c/f_u$ measured at hadron colliders, which have both different production mechanisms and different kinematics. Evaluating $f_c$ at the $Z$ peak with $e^+e^-$ by means of NRQCD mildens the constraint by a factor of $3-4$. A more conservative estimate would further take the theoretical uncertainties into account. 

In light of the above considerations, each NP scenario is analysed under three different assumptions: $\bbc<10,30,60\%$. These constraints are imposed as a hard cut on the region of parameter space allowed for the fit. 

\section{Fit results}

\begin{figure*}[th]
\begin{center}
{
\includegraphics[width=62mm]{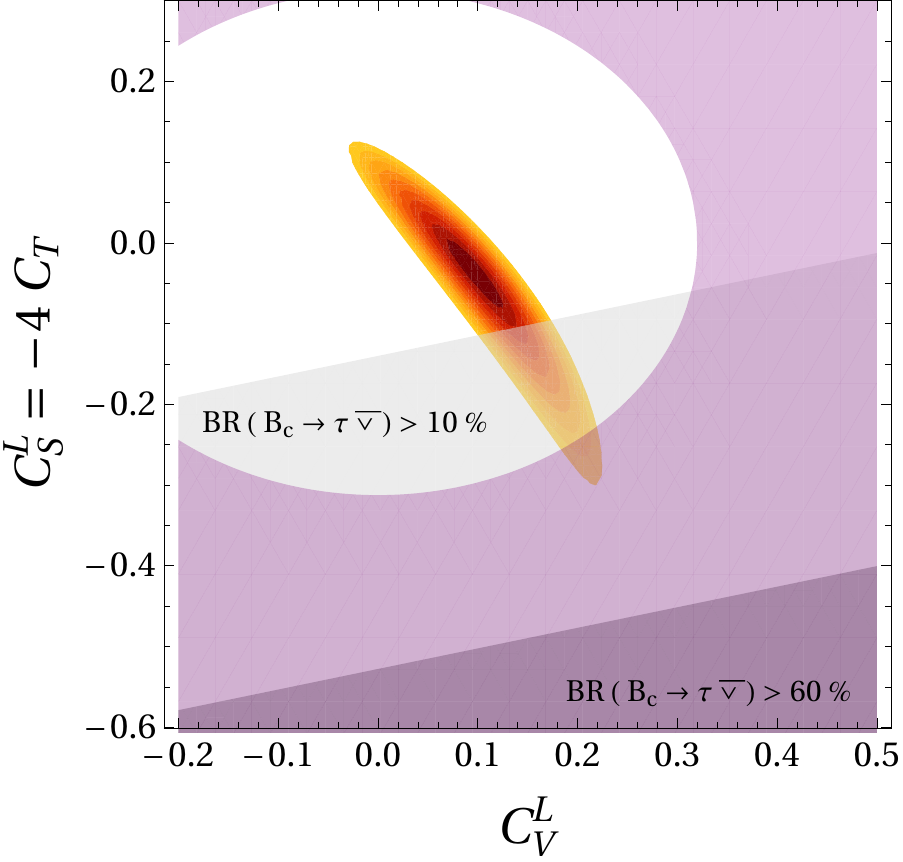}
}
{
\includegraphics[width=73mm]{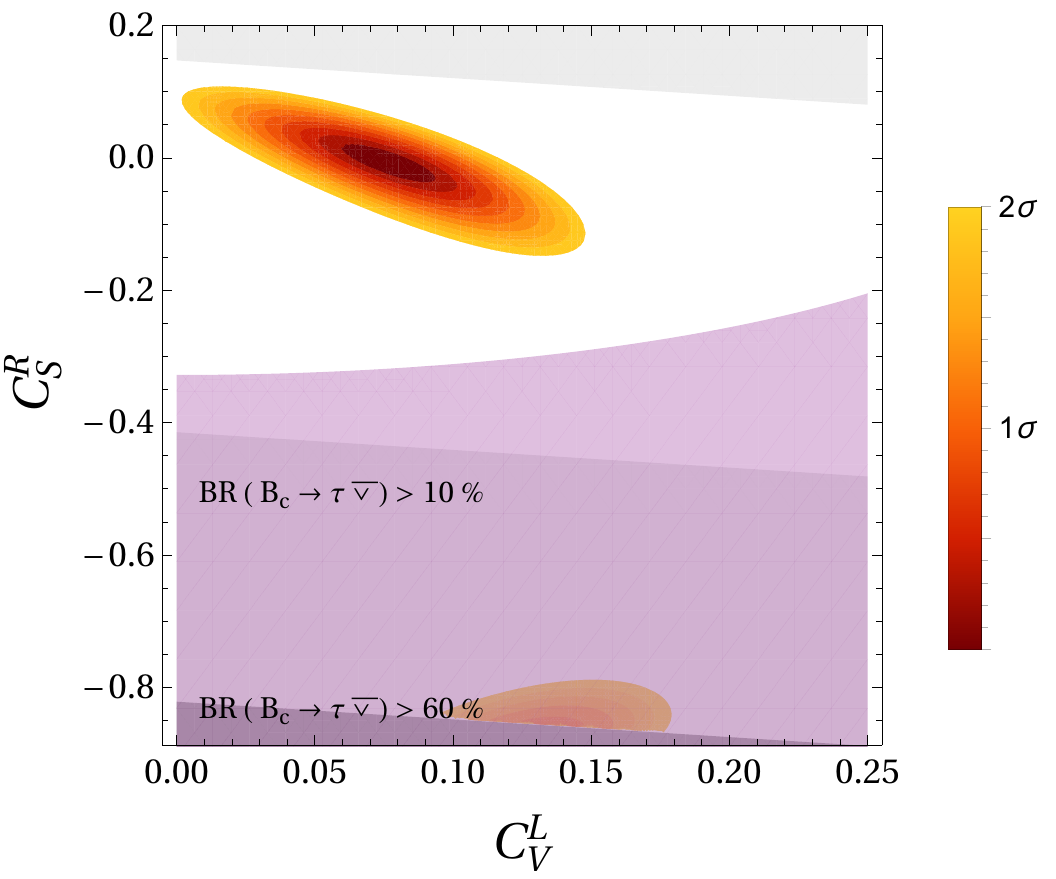}
}
\\
\vspace{0.4cm}
\hspace{0.5cm}
{
\includegraphics[width=60mm]{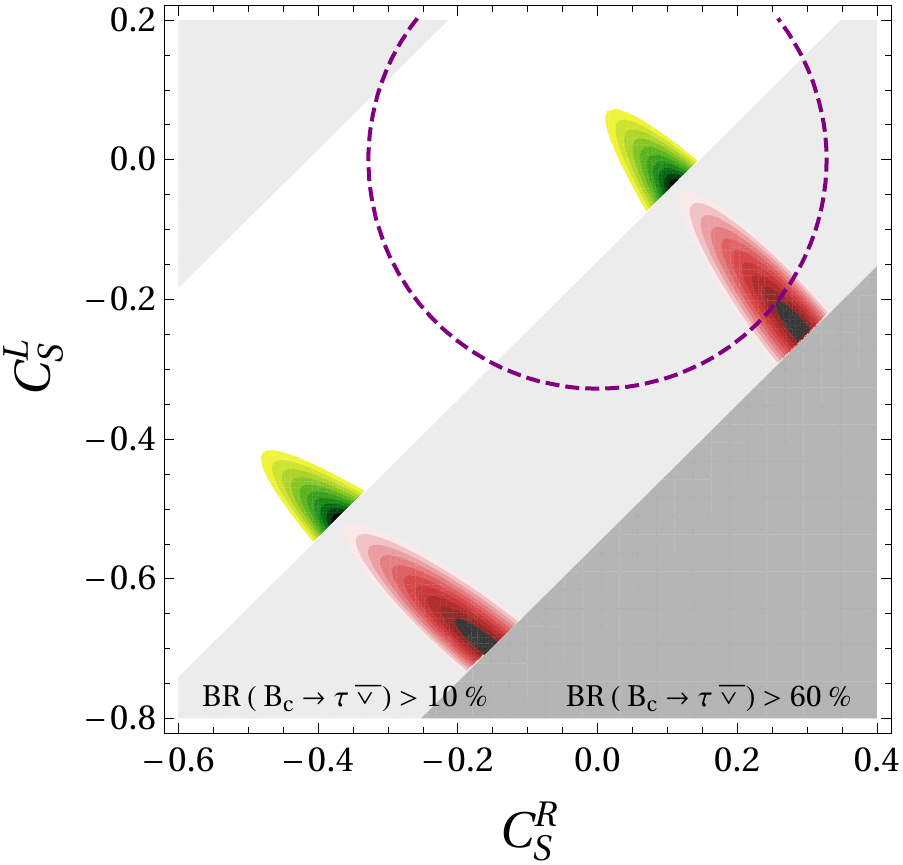} 
}
{
\includegraphics[width=75mm]{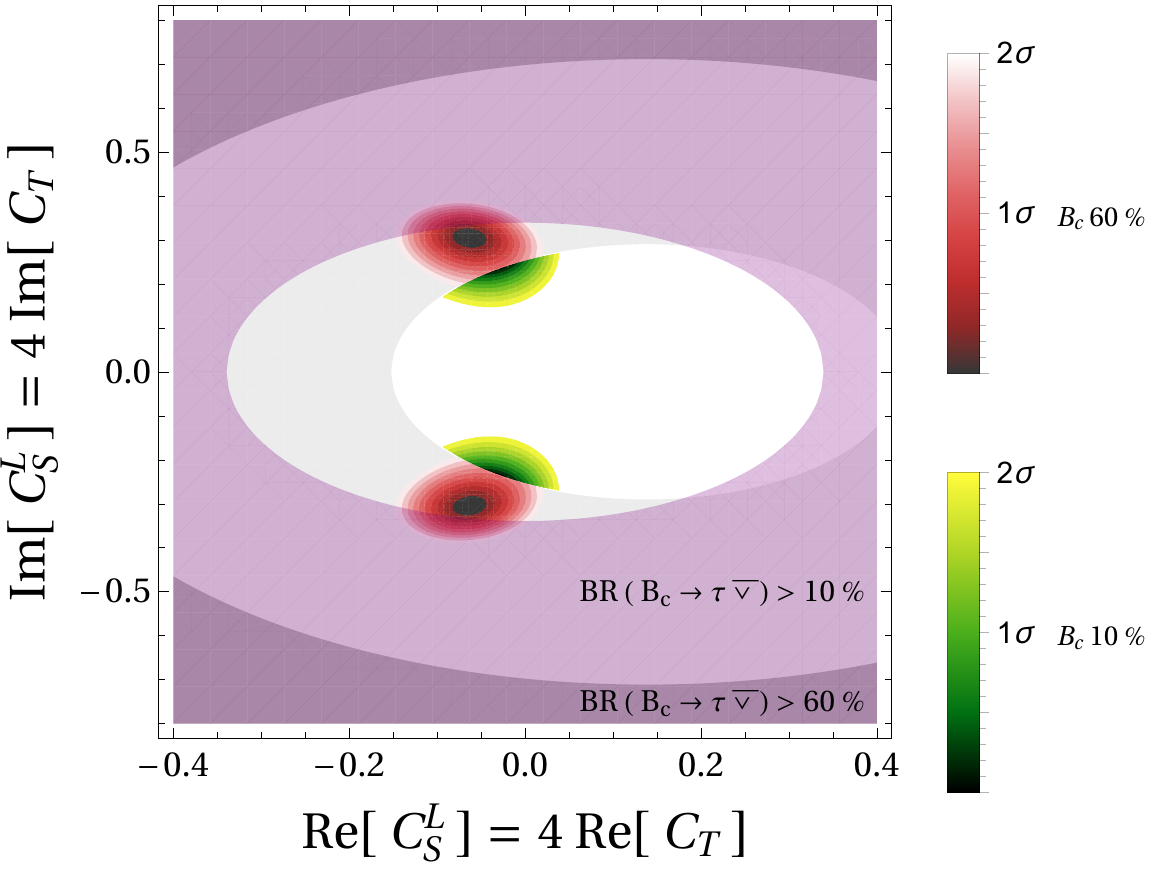}
}
\caption{$2\sigma$ regions of the 2D fit.~\cite{Blanke:2019qrx}}
\label{WCdouble}
\end{center}
\end{figure*}

The results of the fits from~\cite{Blanke:2019qrx} are displayed in Tables~\ref{tab:results1D},~\ref{tab:results2D}. The subscript, where present, refers to the limit on $\bbc$. Its absence indicates that the result does not change when changing the limit on $\bbc$. For each scenario we quote the goodness of fit in terms of $p$-value and the pull of the best-fit point with the SM. The last six columns display the values of the observables at the best-fit point. For the measured ones ($\rd, \rds, \fld, \ptauds$) we also show the pull with respect to the experimental value. The one- and two-$\sigma$ intervals for the 1D fits are displayed in Table~\ref{tab:results1D}, while the same regions for the 2D fits are plotted in Figure~\ref{WCdouble}. The purple regions in scenarios $(C_V^L,\,C_S^L=-4C_T),(C_V^L,\,C_S^R),({\rm Re}[C_S^L=4C_T],\,{\rm Im}[C_S^L=4C_T])$ are excluded at $2\sigma$ by collider bounds~\cite{Greljo:2018tzh}. These constraints are displayed as a dashed line for $(C_S^R,\,C_S^L)$, since a collider study of this scenario requires a model-dependent analysis rather than an EFT one.

\begin{table*}[h]
\begin{center}
\caption{Results of the fit with one-dimensional scenarios.~\cite{Blanke:2019qrx}}
	\begin{tabular}{|c||c|c|c|c|c||c|c|c|c|c|c|c|}\hline
1D hyp.   & best-fit & $1\,\sigma$ range & $2\,\sigma$ range & $p$-value (\%)
& pull$_{\rm SM}$  & ${\cal R}(D)$ & ${\cal R}(D^*)$& $F_L({D^{*}})$ & $P_\tau (D^*)$ & $P_\tau (D)$ & ${\cal R}(\Lambda_c) $\\
\hline\hline    		
$C_V^L$    &  0.07 &   [0.05, 0.09]  & [0.04, 0.11] &44&  4.0   &
\dline{0.347  $+0.2\,\sigma$} &  \dline{0.292  $-0.2\,\sigma$}    &  \dline{0.46  $-1.6\,\sigma$}   & \dline{$-0.49$  $-0.2\,\sigma$}& \dline{0.32 \\}&\dline{0.38 \\}\\
\hline 		
$C_S^R$    & 0.09  &   [0.06, 0.11]   &  [0.03, 0.14]  & 2.7 &  3.1 & \dline{0.380 $+1.4\,\sigma$} & \dline{0.260 $-2.6\,\sigma$}& \dline{0.47 $-1.5\,\sigma$}& \dline{$-0.46$  $-0.1\,\sigma$}& \dline{0.46 \\}&\dline{0.36 \\}\\
\hline  
$C_S^L$    &  0.07 &  [0.04, 0.10]    &  $[-0.00, 0.13]$  &0.26 &  2.1   &  \dline{0.364 $+0.8\,\sigma$}& \dline{0.250 $-3.3\,\sigma$}&   \dline{0.45 $-1.7\,\sigma$}  &\dline{$-0.51$ $-0.2\,\sigma$}& \dline{0.44 \\ }&\dline{0.35 \\ }\\
\hline
\scalebox{.9}{$C_S^L=4C_T$} & $-0.03$  &   [$-0.07$, $0.01$]   &  [$-0.11$, 0.04]  & 0.04& 0.7    & 
\dline{0.278 $-2.1\,\sigma$}& \dline{0.263 $-2.3\,\sigma$}  & \dline{0.46 $-1.6\,\sigma$}     &\dline{$-0.47$ $-0.2\,\sigma$} & \dline{0.27 \\}&\dline{0.33 \\}\\
\hline
\end{tabular}
	\label{tab:results1D}
\end{center}
\end{table*}

\begin{table*}[h]
\begin{center}
 \caption{Results of the fits with two-dimensional scenarios.~\cite{Blanke:2019qrx}}
\begin{tabular}{|c||c|c|c||c|c|c|c|c|c|c|c|c|}\hline
  	2D hyp.  & best-fit& $p$-value {(\%)} & pull$_{\rm SM}$  & ${\cal R}(D)$ & ${\cal R}(D^*)$ & $F_L({D^{*}})$ &  $P_\tau (D^*)$ & $P_\tau (D)$ & ${\cal R}(\Lambda_c) $\\\hline\hline
  	$(C_V^L,\,C_S^L=-4C_T)$&($0.10,-0.04)$&29.8&3.6&
  	\dline{0.333 $-0.2\,\sigma$}& \dline{0.297 $+0.2\,\sigma$}&\dline{0.47 $-1.5\,\sigma$} &\dline{$-0.48$ $-0.2\,\sigma$}& \dlinem{0.25 \\}& \dlinem{0.38 \\}\\ \hline
  	$\left(C_S^R,\,C_S^L)\right|_{60 \%}$& \parbox{7em}{\rule{0pt}{2.3ex}($0.29,-0.25)$ $(-0.16,-0.69)$\strut}&75.7&3.9&
  	\dline{0.338 $~~~0.1\,\sigma$} &\dline{0.297 $+0.1\,\sigma$}& \dline{0.54 $-0.7\,\sigma$}&\dline{$-0.27$ $+0.2\,\sigma$}& \dlinem{0.39 \\}&\dlinem{0.38 \\}\\ \hline
  	$\left(C_S^R,\,C_S^L)\right|_{30 \%}$&\parbox{7em}{\rule{0pt}{2.3ex}($0.21,-0.15)$ $(-0.26,-0.61)$\strut}&30.9&3.6&
  	\dline{0.353 $+0.4\,\sigma$}&\dline{0.280 $-1.1\,\sigma$}&\dline{0.51  $-1.0 \,\sigma$}& \dline{$-0.35$ $~~~0.0\,\sigma$}&\dlinem{0.42 \\}&\dlinem{0.37\\}\\ \hline
  	$\left(C_S^R,\,C_S^L)\right|_{10 \%}$
  	&\parbox{7em}{\rule{0pt}{2.3ex}($0.11,-0.04)$ $(-0.37,-0.51)$\strut}&2.6&2.9&\dline{0.366 $+0.9\,\sigma$} &\dline{0.263 $-2.3\,\sigma$}&\dline{0.48 $-1.4\,\sigma$} &\dline{$-0.44$ $-0.1\,\sigma$} &\dlinem{0.44 \\}&\dlinem{0.36\\}\\ \hline 
  	$(C_V^L,\,C_S^R)$&($0.08,-0.01)$&26.6&3.6&
  	\dline{0.343 $+0.1\,\sigma$}&\dline{0.294 $-0.1\,\sigma$}&\dline{0.46 $-1.6\,\sigma$} &\dline{$-0.49$ $-0.2\,\sigma$}&\dlinem{0.31 \\}&\dlinem{0.38 \\}\\ \hline 
  	\scalebox{.86}{$\left.({\rm Re}[C_S^L=4C_T],\,{\rm Im}[C_S^L=4C_T])\right|_{60,30\%}$}&($-0.06,\pm 0.31)$&25.0&3.6&
  	\dline{0.339 $0.0\,\sigma$}&\dline{0.295 0.0 $\,\sigma$}
  	&\dline{0.45 $-1.7\,\sigma$} 
  	&\dline{$-0.41$ $-0.1\,\sigma$}&\dlinem{0.41 \\}&\dlinem{0.38 \\}
  	\\ \hline
  	\scalebox{.87}{$\left. ({\rm Re}[C_S^L=4C_T],\,{\rm Im}[C_S^L=4C_T])\right|_{10\%}$} &($-0.03,\pm 0.24)$&5.9&3.2&
  	\dline{0.330 $-0.3\,\sigma$}& \dline{0.275 $-1.4\,\sigma$}&\dline{0.46  $-1.6\,\sigma$}& \dline{$-0.45$ $-0.1\,\sigma$} &\dlinem{0.38 \\}&\dlinem{0.36 \\}\\ \hline
\end{tabular}
  	\label{tab:results2D}
  	\end{center}
  \end{table*}

Concerning $\bbc$, the most striking result from Table~\ref{tab:results2D} is that with a $60\%$ limit, the scenario $(C_S^R,\,C_S^L)$ is the one preferred by the current experimental data. Its $p$-value diminishes drastically as soon as we impose a more severe $\bbc$ constraint. We conclude that a description of the $\rdrds$ anomaly in terms of charged Higgs predicts $\bbc>30\%$.

\subsection{Correlations between observables and $\boldmath{\rlam}$ sum rule}
\begin{figure*}[th]
\begin{center}
{\includegraphics[width=67mm]{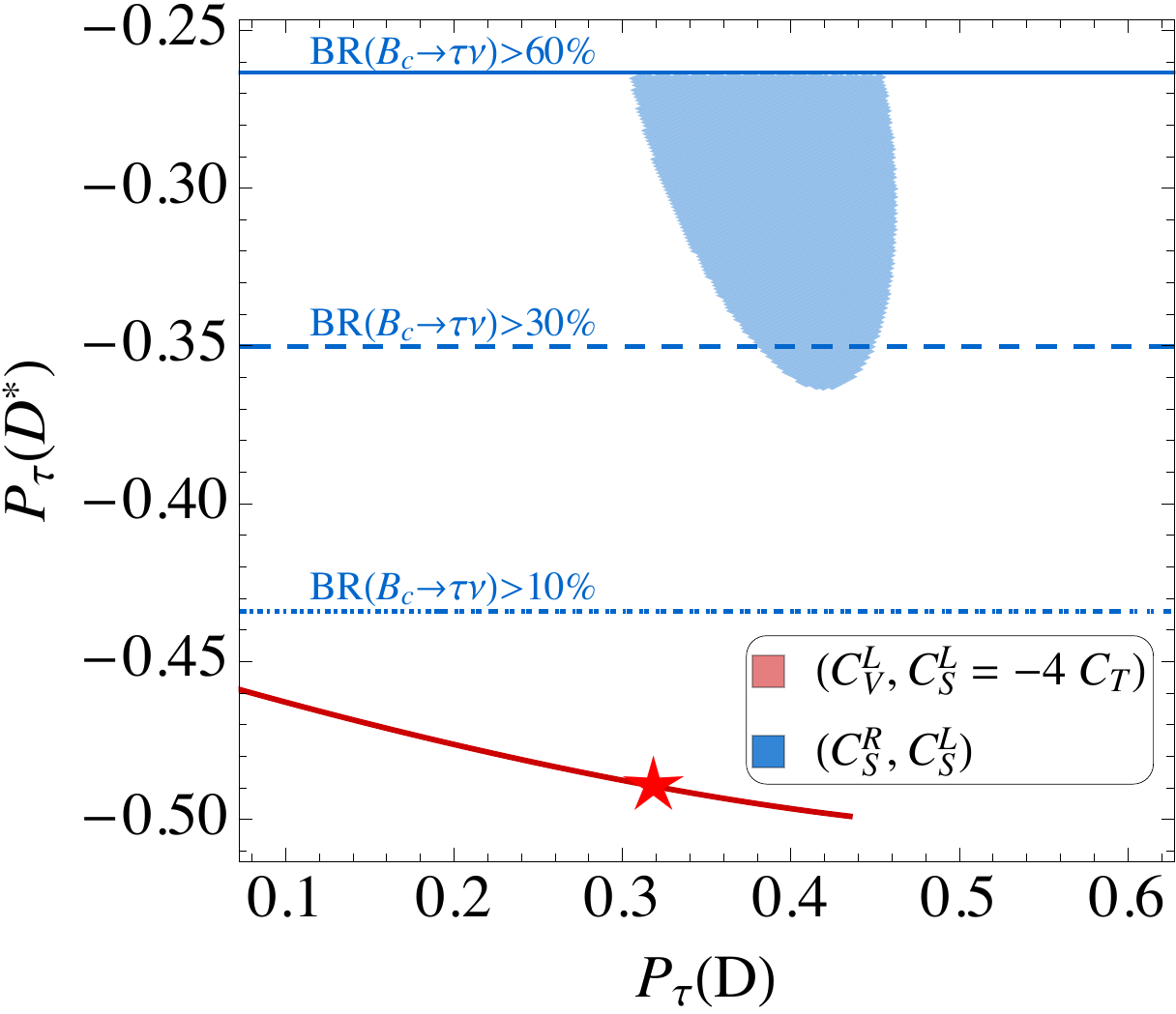}}
{\includegraphics[width=67mm]{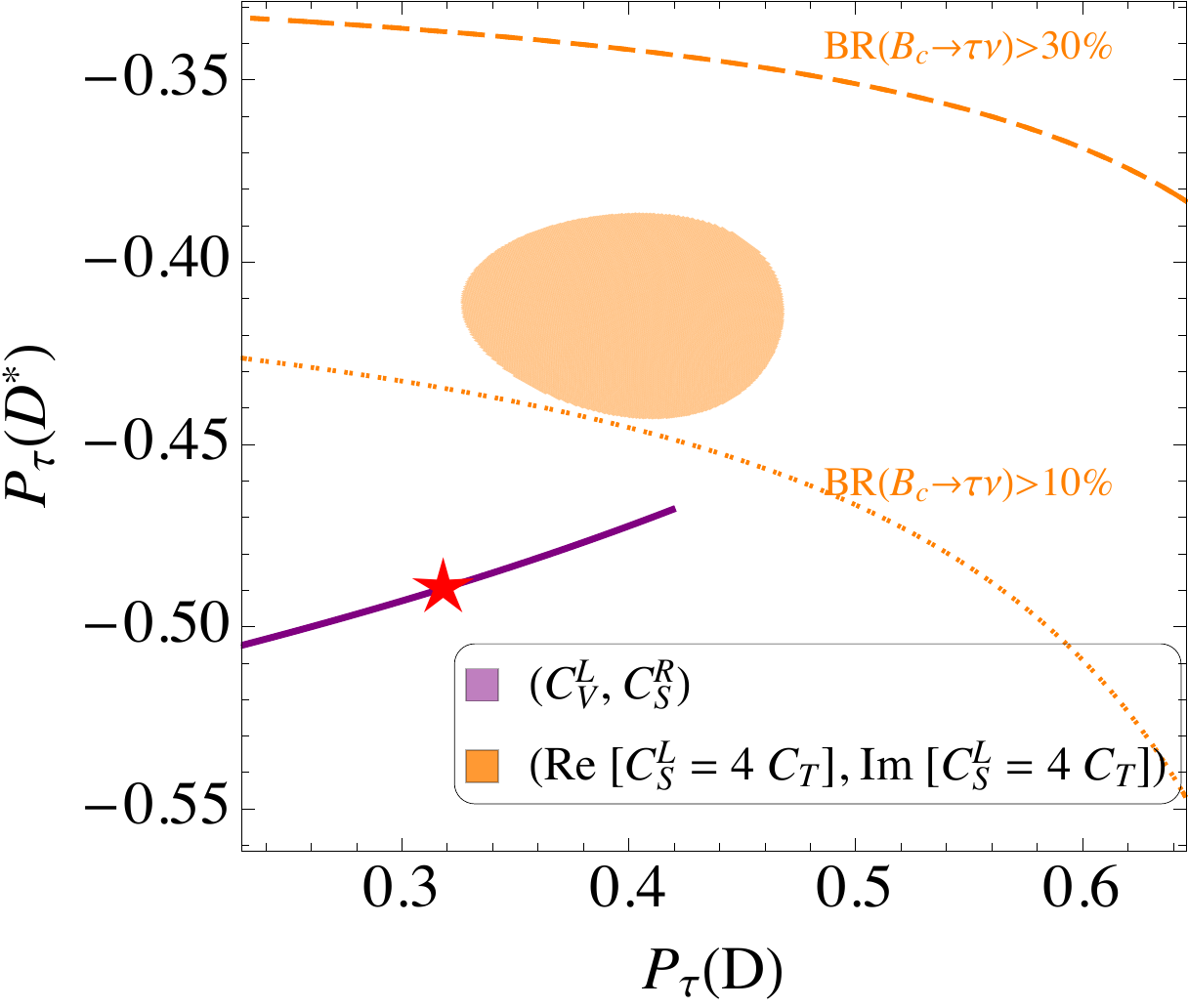}}
\\
\vspace{0.4cm}
\hspace{0.5cm}
{\includegraphics[width=67mm]{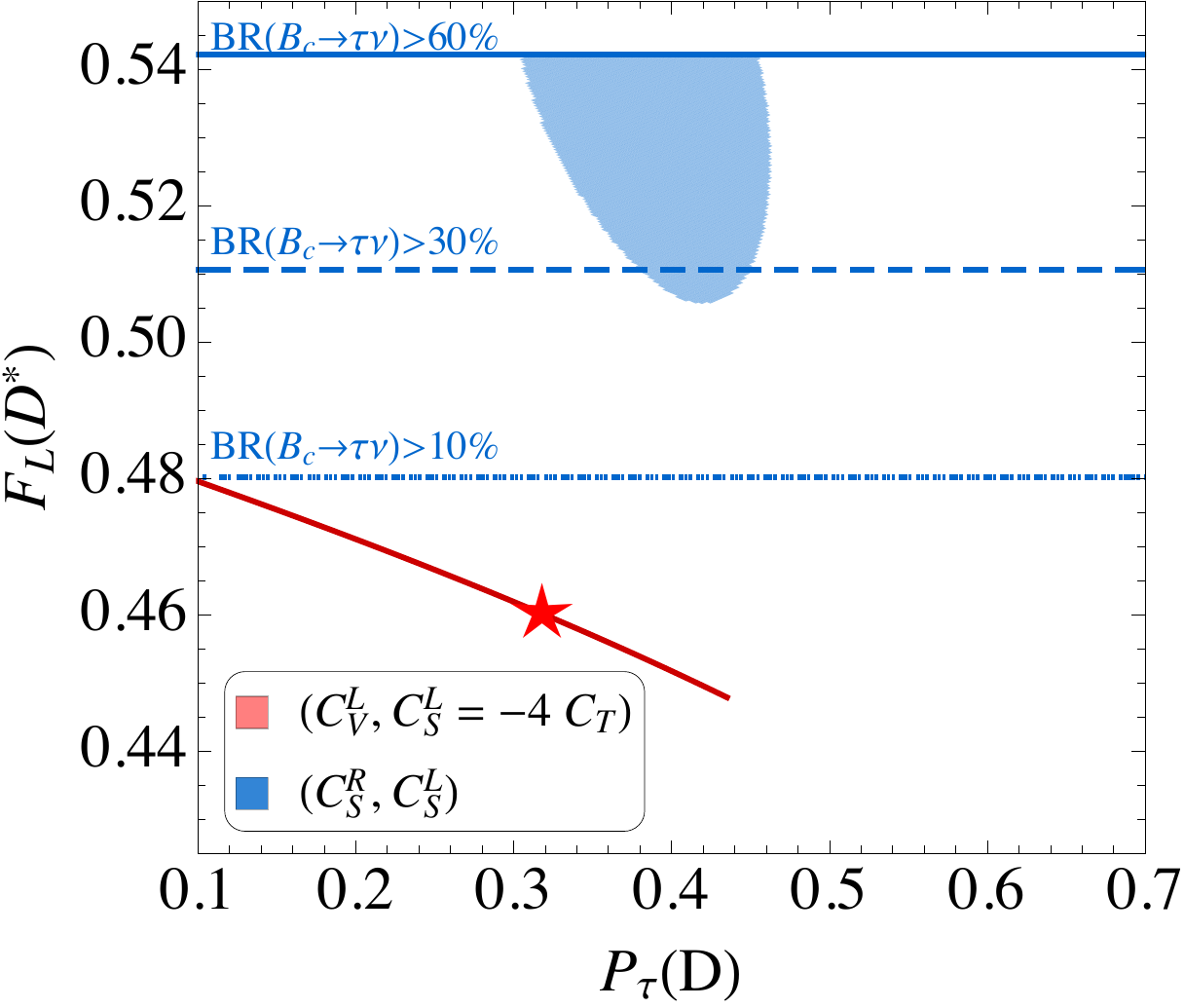} }
{\includegraphics[width=67mm]{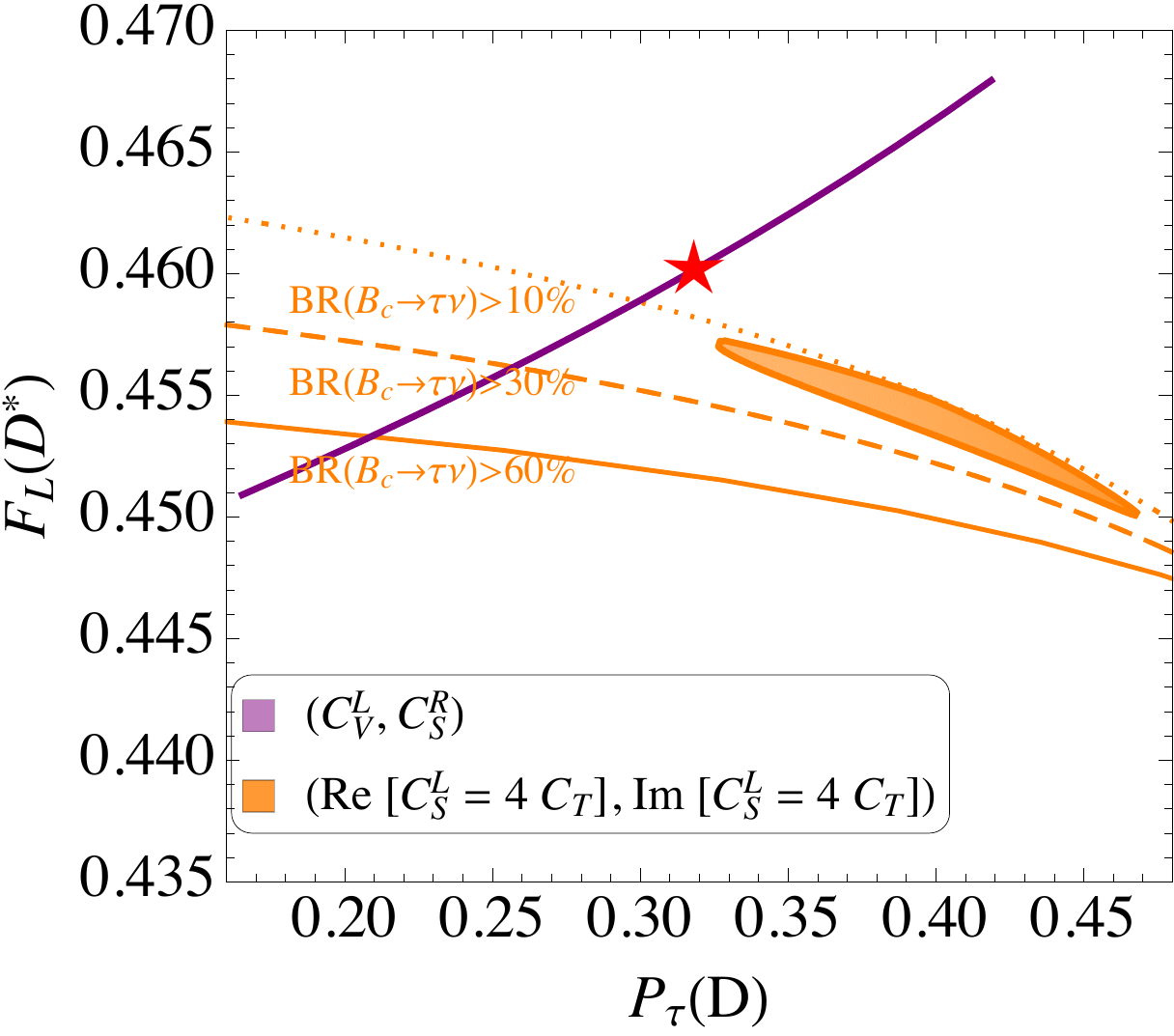}}
\\
\vspace{0.4cm}
\hspace{0.5cm}
{\includegraphics[width=67mm]{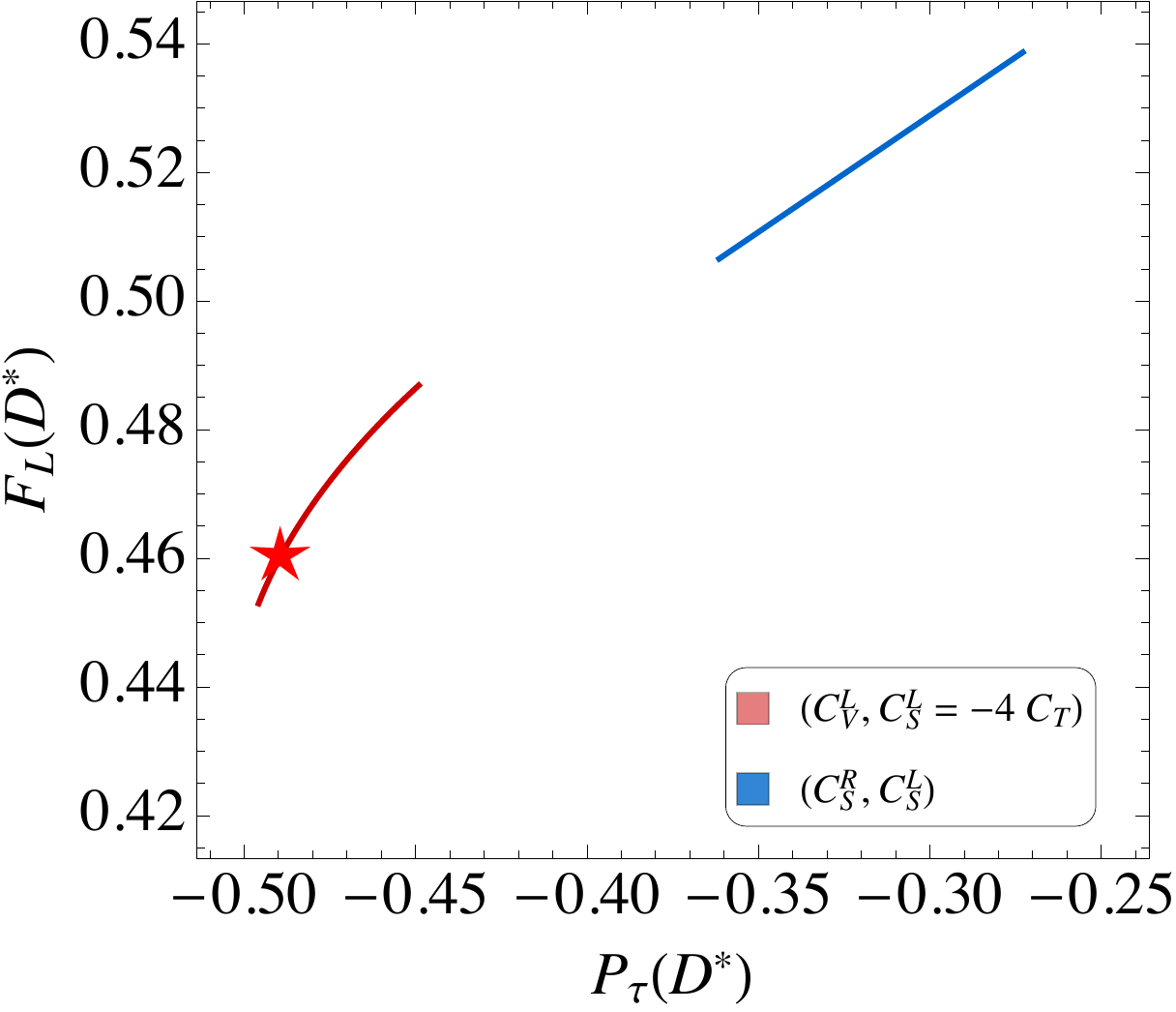} }
{\includegraphics[width=67mm]{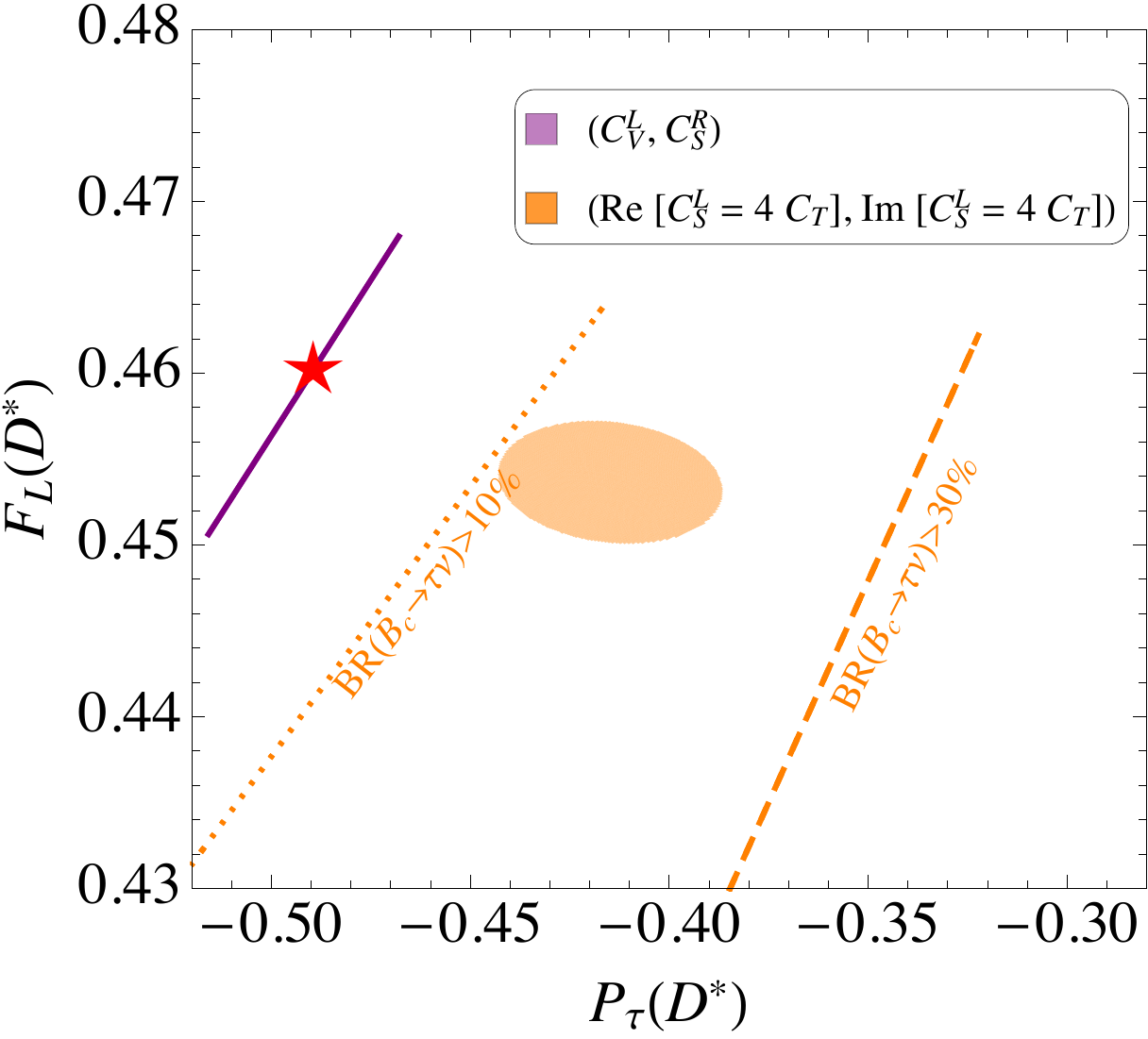}}
\caption{Correlation plots among polarisation observables for the 2D fits~\cite{Blanke:2019qrx}. The red star represents the Standard Model prediction.}
\label{fig:correlang}
\end{center}
\end{figure*}

\begin{figure*}[th]
\begin{center}
{\includegraphics[width=67mm]{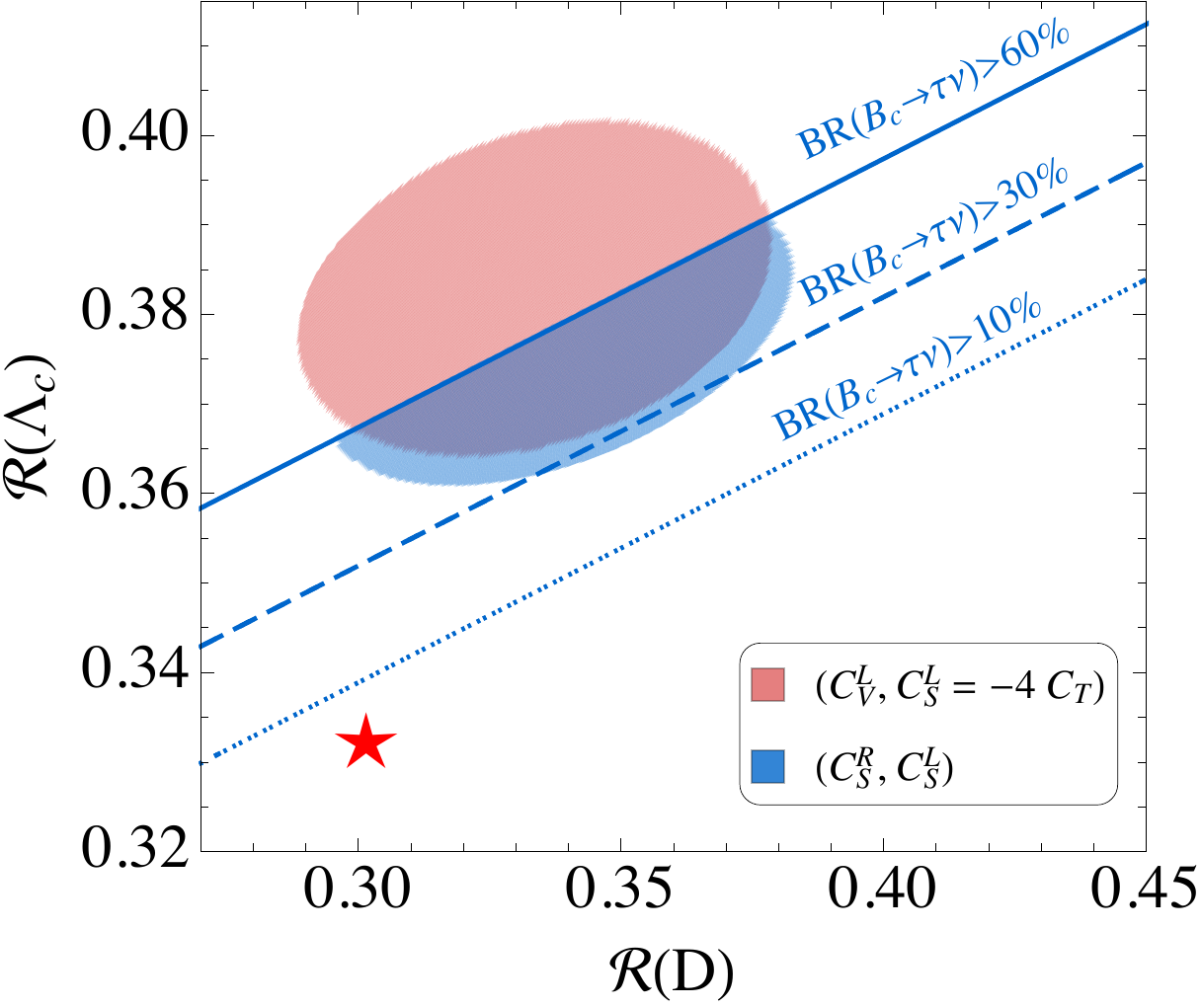} }
{\includegraphics[width=67mm]{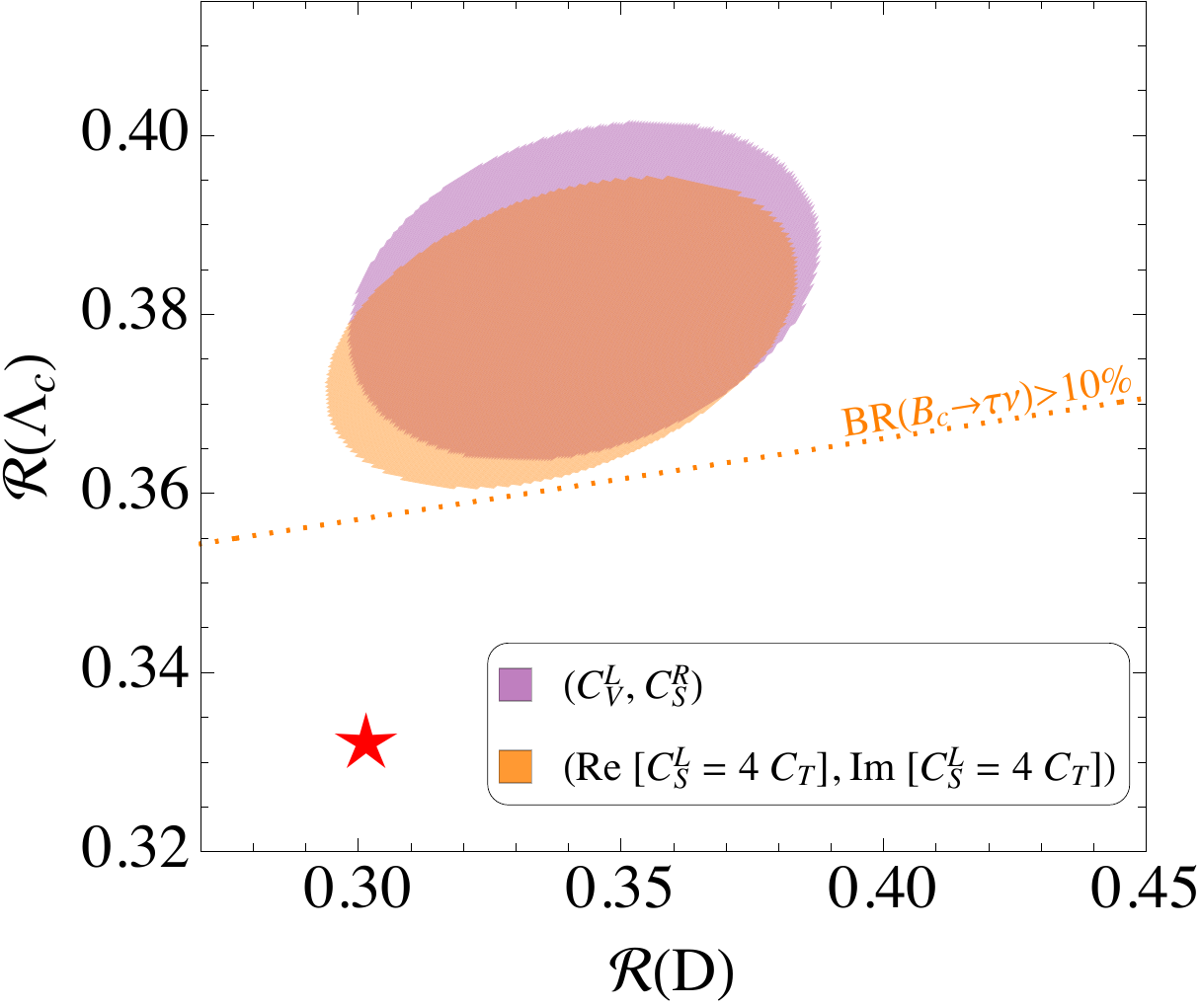}}
\\
\vspace{0.4cm}
\hspace{0.5cm}
{\includegraphics[width=67mm]{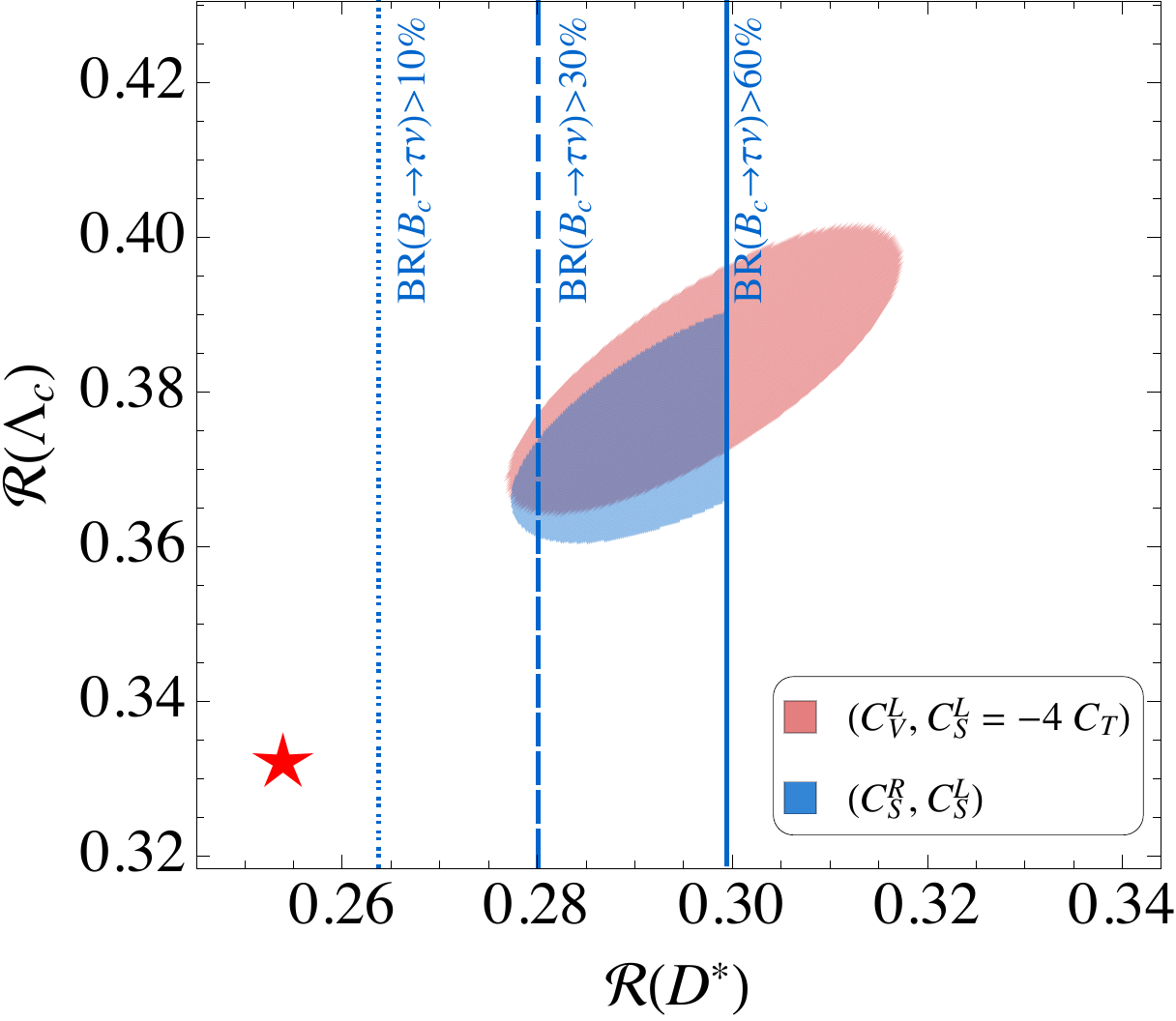} }
{\includegraphics[width=67mm]{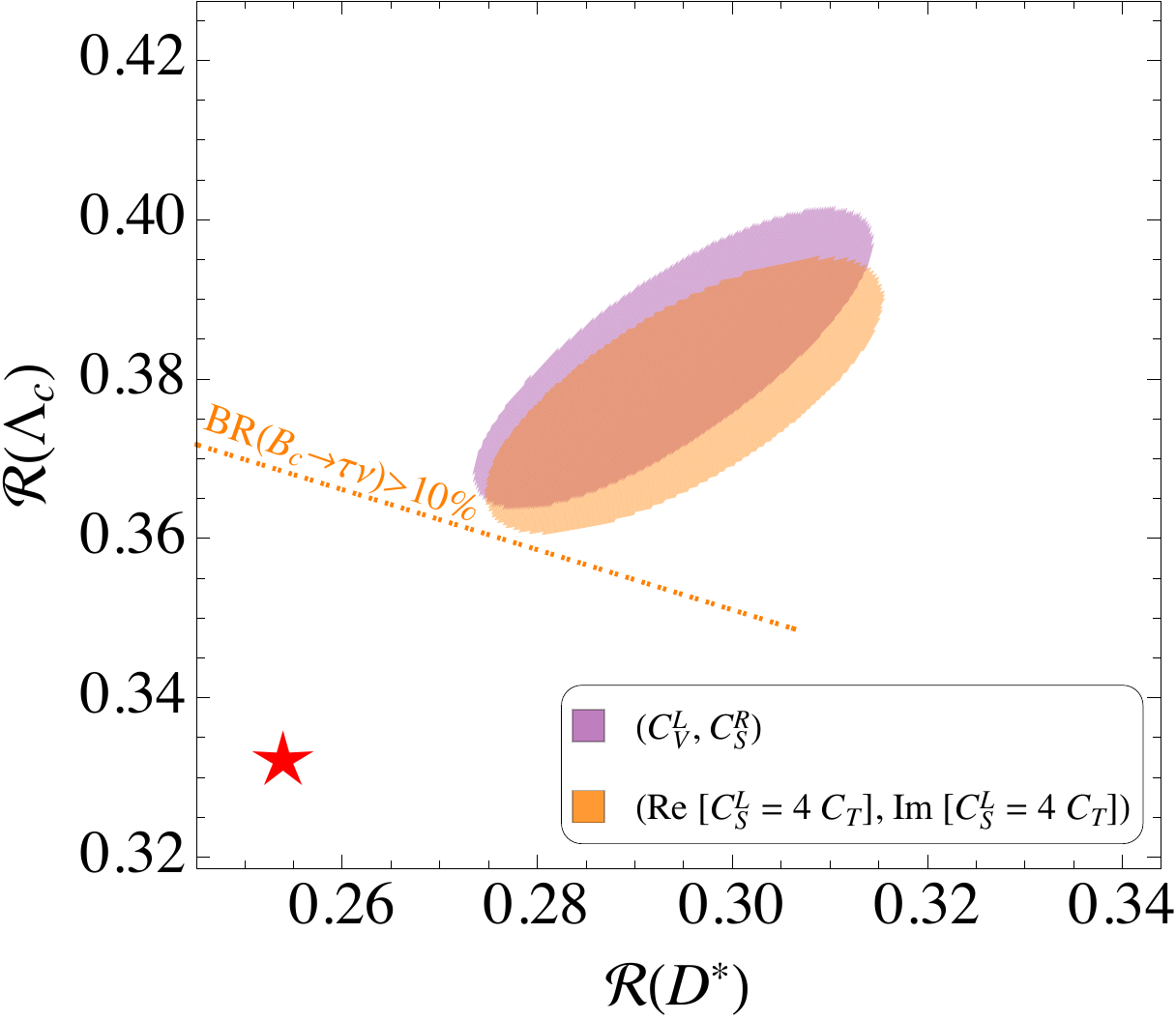}}
\caption{Correlation plots between $\rlam$ and $\rdrds$ for the 2D fits~\cite{Blanke:2019qrx}. The red star represents the Standard Model prediction.}
\label{fig:correllam}
\end{center}
\end{figure*}

For the two dimensional scenario we also analysed the correlation between the observables in the last six columns of Table~\ref{tab:results2D}. In order to do so, we projected the two-sigma regions resulting from the fits with the $\bbc<60\%$ limit into planes having as axes two out of the six observables. These plots are displayed in Figures~\ref{fig:correlang}~and~\ref{fig:correllam} and allow us to draw two conclusions. 

From Figure~\ref{fig:correlang} we see that in planes in which one of the axes is a polarisation observable, the sigma regions of different scenarios separate clearly, hence indicating that these observables have a strong impact in distinguishing among models. In particular, a closer look at Table~\ref{tab:results2D} reveals that the recent measurement of $\fld$ favours the scenario $(C_S^R,\,C_S^L)$.

In Figure~\ref{fig:correllam}, instead, we see that the value of $\rlam$ predicted in models fitting $\rdrds$ is always increased with respect to its SM prediction~\cite{Detmold:2015aaa,Bernlochner:2018bfn}. This enhancement can be traced back to the sum rule
\begin{eqnarray}
\frac{\mathcal{R}(\Lambda_c)}{\mathcal{R}_{\rm SM}(\Lambda_c)}
  \simeq 0.262 \frac{\mathcal{R}(D)}{\mathcal{R}_{\rm SM}(D)} + 0.738 \frac{\mathcal{R}(D^*)}{\mathcal{R}_{\rm SM}(D^*)},  \label{eq:sumrule}
\end{eqnarray}
which holds irrespective of the NP model considered, and that can be understood in the heavy-quark limit. Substituting the current experimental averages of $\rdrds$, we find 
\begin{equation}
 \mathcal{R}(\Lambda_c) = 0.38 \pm 0.01 \pm 0.01      \label{eq:predlc2}\,,
\end{equation}
where the first error arises from the experimental uncertainty of $\rdrds$, 
while the second comes from the form factors.

\section{Summary}
Motivated by the $\rdrds$ anomaly, we updated the fit of $b\to c\tau\nu$ data, including the recent experimental results from the Belle collaboration and restricting to scenarios with a single additional mediator. We revised the limit from $\bbc$ and analysed its impact on each of the scenarios. The fit allowed us to appreciate the model-resolving power of polarisation observables, and to conclude that if the origin of the $\rdrds$ anomaly is new physics, we expect a value of $\rlam$ higher than the one predicted by the Standard Model, irrespective of which additional particle mediates the decay.
\begin{acknowledgments}
I am grateful to Monika Blanke, Andreas Crivellin, Stefan de Boer, Teppei Kitahara, Uli Nierste and Ivan Ni\v{s}and\v{z}i\'c for the fruitful collaboration that led to the results presented in this proceeding. I would also like to thank the organisers of the FPCP 2019 and to acknowledge the support of the DFG-funded Doctoral School KSETA and of the research training group GRK 1694.
\end{acknowledgments}

\bigskip 

\end{document}